\documentclass{nature}

\usepackage{graphicx}
\usepackage{amsmath}
\usepackage{amssymb}
\usepackage{textcomp}
\usepackage{gensymb}
\usepackage{xcolor}
\usepackage{lineno}
\usepackage{soul}
\usepackage{colortbl}
\usepackage[labelfont=bf]{caption}
\usepackage{multibib}
\usepackage{float}
\usepackage{soul}

\makeatletter
\let\saved@includegraphics\includegraphics
\AtBeginDocument{\let\includegraphics\saved@includegraphics}
\renewenvironment*{figure}{\@float{figure}}{\end@float}
\makeatother

\newcommand{\reftextit}[1]{}
\title{Long-lived Zone-boundary Magnons in an Antiferromagnet}

\author{Jeongheon Choe$^{1,2,\dagger}$,
David Lujan$^{1,2,\dagger}$,  Gaihua Ye$^{3}$,  Cynthia Nnokwe$^{3}$,
Bowen Ma$^1$,  Jiaming He$^{4}$, 
Frank Y. Gao$^{1,2}$, 
T. Nathan Nunley$^{1,2}$,
Aritz Leonardo$^{5}$,
Mikel Arruabarrena$^{6}$,
Andres Ayuela$^{6}$,
Jianshi Zhou$^{2,4}$,
Martin Rodriguez-Vega$^{1,2,7,*}$
Gregory A. Fiete$^{7,8,9}$, 
Rui He$^{3,*}$\& 
Xiaoqin Li$^{1,2,*}$}

\begin{document} 

\maketitle 

{\renewcommand{\baselinestretch}{1.5}
\begin{affiliations}
    \item Department of Physics and Center for Complex Quantum Systems, The University of Texas at Austin, Austin, Texas, 78712, USA.
    \item Center for Dynamics and Control of Materials, The University of Texas at Austin, Austin, Texas, 78712, USA
    \item Department of Electrical and Computer Engineering, Texas Tech University, Lubbock, Texas, 79409, USA 
    \item Department of Mechanical Engineering, The University of Texas at Austin, Austin, Texas, 78712, USA
    \item EHU Quantum Center, University of the Basque Country UPV/EHU, Bilbao, Spain
    \item Material Physics Center (CFM), Donostia 20018, Spain
    \item Department of Physics, Northeastern University, Boston, Massachusetts, 02115, USA
    \item Quantum Materials and Sensing Institute, Northeastern University, Burlington, MA, 01803 USA
    \item Department of Physics, Massachusetts Institute of Technology, Cambridge, Massachusetts, 02139, USA
  
    \thanks{$^{\dagger}$ These authors contributed equally to this work.}\\
    $^{*}$\thanks{Corresponding authors:\textcolor{blue}{{{elaineli@physics.utexas.edu}; rui.he@ttu.edu}; rodriguezvega.physics@gmail.com}}
\end{affiliations}
}
\newpage
{\renewcommand{\baselinestretch}{1.7}

\begin{abstract}
Antiferromagnetic (AFM) insulators
exhibit many desirable features for spintronic applications such as fast dynamics in the THz range and robustness to fluctuating external fields. However, large damping typically associated with THz magnons presents a serious challenge for THz magnonic applications. Here, we report long-lived short-wavelength zone boundary magnons in the honeycomb AFM insulator CoTiO$_3$, recently found to host topological magnons. We find that its zone-boundary THz magnons exhibit longer lifetimes than its zone-center magnons. This unusual momentum-dependent long magnon lifetime originates from several factors including the antiferromagnetic order, exchange anisotropy, a finite magnon gap, and magnon band dispersion. Our work suggests that magnon-magnon interaction may not be detrimental to magnon lifetimes and should be included in future searches for topological magnons.
\end{abstract}

}

\maketitle
\renewcommand{\baselinestretch}{2}
\setlength{\parskip}{7pt}

Topological magnons offers an enticing solution to the outstanding challenge of damping associated with THz magnonics~\cite{mcclarty2021topological}. As newly identified bosonic excitations in  crystalline magnetic insulators, topological magnons possess distinct characteristics in comparison to their electronic counterpart~\cite{mcclarty2021topological,mook2021interaction,PhysRevX.8.011010}. On the one hand, surface or edge states of topological magnons are protected from local defects in materials~\cite{PhysRevB.87.144101,PhysRevB.95.014435}, analogous to other topological systems (e.g. electrons, acoustic phonons, and photonic crystals). On the other hand, 
magnons only exist as excited states and do not obey particle number conservation. Consequently, only long-lived magnons, regardless of their band topology, can offer the benefit associated with topologically protected states. 

While many materials have been predicted to host topological magnons, a simple relation between magnon topology (properties of single-particle bands) and their lifetime does not exist. A prediction of the magnon lifetime must take into account magnon-magnon interactions, which necessitates theories that go beyond single-particle band descriptions~\cite{chernyshev2016damped,koyama2023flavor,habel2024breakdown}. For fermions, only electrons near the Fermi level participate in the excitation processes, allowing one to focus on low-energy Dirac cones.  In contrast, bosons, unrestricted by the Pauli exclusion principle, require consideration of the entire Brillouin zone. Notably, magnons residing near the saddle points of a dispersion relation may play a particularly significant role in both electrically and optically detected magnon signals due to their high density of states. In the presence of magnon-magnon scattering, overlapping single-magnon bands with two- or three-magnon continua results in magnon damping.~\cite{bayrakci2006spin} Interestingly, in chiral honeycomb ferromagnets, theoretical studies have further identified specific conditions (e.g. external magnetic field and temperature) under which topological phase transitions occur in the presence of magnon interactions~\cite{mook2021interaction,PhysRevX.8.011010}. Recent experiments based on inelastic neutron scattering~(INS)~\cite{chen2018topological,bao2018discovery} and the thermal Hall effect~\cite{onose2010observation,zhang2021anomalous} have identified several classes of promising materials. However, direct detection of in-gap surface states or images of topologically protected edge states similar to experiments on electronic topology are still out of reach~\cite{chen2009experimental,drozdov2014one}.  

Here, we study CoTiO$_3$ as a model honeycomb AFM insulator with confirmed Dirac magnon bands.~\cite{Yuan2020,elliot2021order} Using polarization-resolved Raman spectroscopy, we identify multiple magnon peaks via temperature- and magnetic-field-dependent experiments. The saddle points at the high symmetry M points on the Brillouin zone boundary lead to an exceptionally large density of states, thus, dominating two-magnon scattering processes. Most intriguingly, these exchange magnons with nanometer wavelengths exhibit a significantly sharper linewidth (limited by instrument resolution) than the $\Gamma$ point magnon at the zone center. This unusual momentum dependence of magnon lifetimes in CTO originates from several combined factors, including the antiferromagnetic order, particle number non-conserving interaction, a small joint two-magnon density of states, exchange anisotropy and single magnon band features. The magnon lifetime decreases in the presence of an external magnetic field as additional three-magnon decay channels emerge in a non-collinear spin configuration. Our finding establishes new guiding principles for identifying antiferromagnetic insulators for spintronic applications, such as terahertz oscillators and magnon-phonon coupling-based cavity.

\begin{figure}[H]
    \centering
    \includegraphics[width = 16 cm]{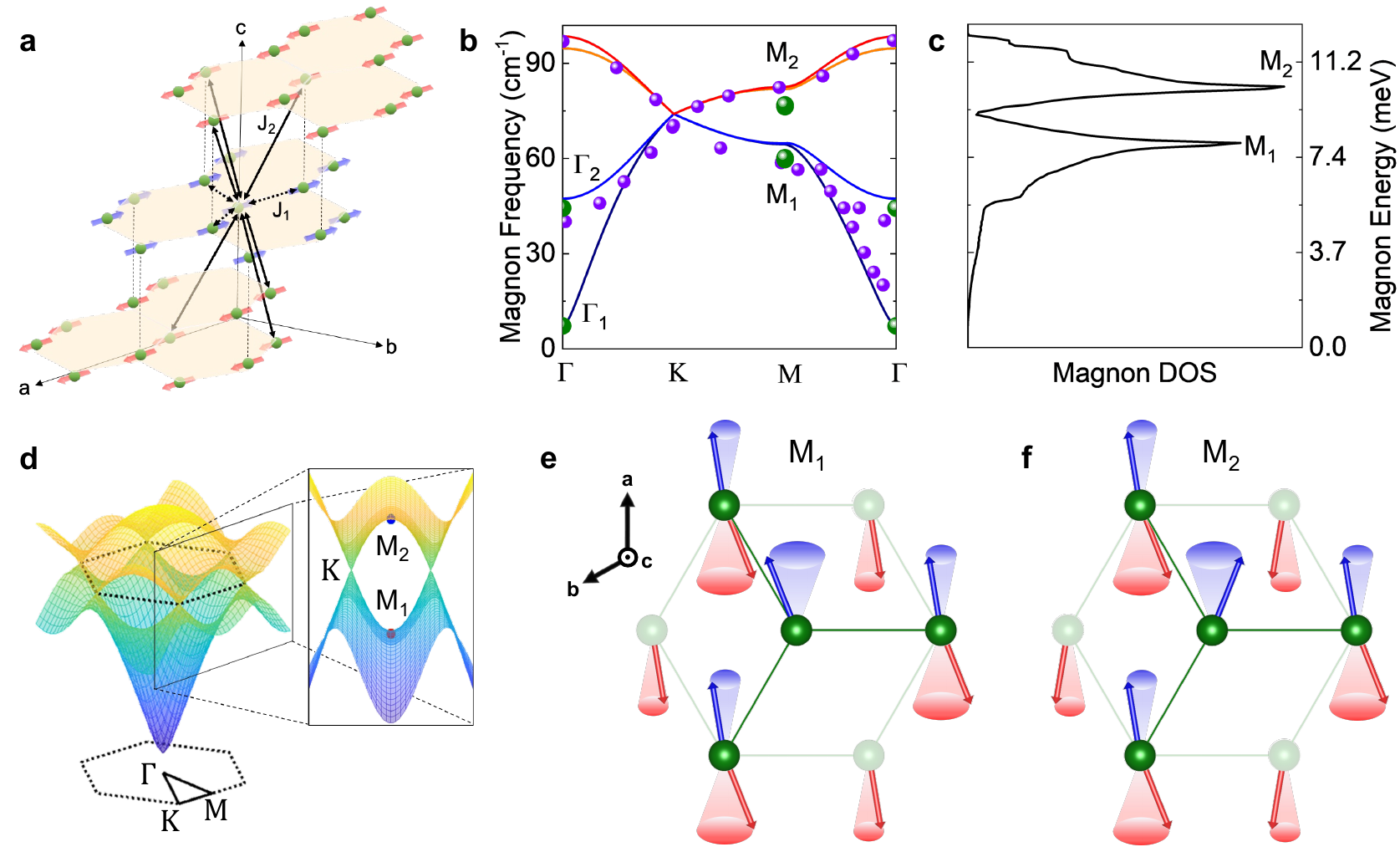}
    \caption{\textbf{Magnons in CoTiO$_3$.} (a) Spin lattice structure of the buckled honeycomb of Co$^{2+}$ ions in the AFM phase. The spins are ordered ferromagnetically (antiferromagnetically) between intralayer (interlayer) spins coupled by the exchange interaction $\textit{J}_{1}$ ($\textit{J}_{2}$). The colored arrows indicate the direction of magnetic moments aligned in the $ab$ plane. (b) Calculated magnon dispersion (color solid lines) along the lines of $\Gamma$-K-M-$\Gamma$ in the Brillouin zone. The green and purple dots are experimental data from our Raman measurements and inelastic neutron scattering (INS) from Ref~\cite{Yuan2020}, respectively. (c) Magnon density of states (DOS) calculated from the magnon dispersion. (d) Schematics of magnon dispersion that highlights a linear dispersion at the $K$ point and saddle-points at M$_1$ (red dot) and M$_2$ (blue dot) points. The hexagon denotes the zone boundary labelled with high symmetry points. (e, f) Schematic of magnon precession for M$_1$ and M$_2$ modes involving two adjacent honeycomb lattices in the $ab$ plane. Atoms in the top (bottom) layer are represented by dark (light) green balls and the arrows describe the phases of the precession. Three atoms (with both blue and red cones) in the two layers overlap in this top view along the $c$-axis.}
    \label{fig:fig1}
\end{figure}

The magnetic structure of CoTiO$_3$ is illustrated in Fig.~\ref{fig:fig1}a. The magnetic moments are ordered ferromagnetically within each plane and coupled antiferromagnetically along the $c$-axis. The Co$^{2+}$ ions are arranged in two-dimensional honeycomb lattices that are slightly buckled. They are stacked along the $c$-axis in an ABC sequence with neighboring planes displaced diagonally by a third of the unit cell. The topological magnon dispersions are reproduced  in Fig.~\ref{fig:fig1}b using a model Hamiltonian $H =\sum_{i\delta} \textit{J}_1 (S^{x}_{i} S^{x}_{i+\delta}  +S^{y}_{i}S^{y}_{i+\delta}) + \sum_{i\gamma}\textit{J}_2 (S_{i} \cdot S_{i+\gamma})$ with exchange couplings $J_1$ = -4.4 meV (XY interaction) and  $J_2$ = 0.57 meV (isotropic Heisenberg interaction) between pairs of Co$^{2+}$ ions that are labeled by the arrows $\textit{J}_1$ and $\textit{J}_2$ in Fig.~\ref{fig:fig1}a. The coupling parameters are extracted from inelastic neutron scattering (INS) experiments\cite{Yuan2020,elliot2021order} (reproduced as purple spheres in Fig.~\ref{fig:fig1}b) and suggest a strong easy-plane anisotropy. The Raman measurements (to be presented later) are overlaid as green spheres.  The comparison to  neutron scattering experiments serves as an important guidance in assigning the magnon modes in our Raman spectra.  

The magnon density of states (DOS) calculated from the dispersion is shown in Fig.~\ref{fig:fig1}c. At the M points, the DOS is much larger than other high symmetry points in the Brillouin zone, as a consequence of the saddle shape of magnon bands at the M points as illustrated in Fig.~\ref{fig:fig1}d. The M$_1$ (M$_2$) mode corresponds to an acoustic (optical) mode with in-phase (out-of-phase) precession in a semi-classical picture as illustrated in Fig.~\ref{fig:fig1}e,f. 
The magnetic moments in adjacent layers are oppositely aligned and undergo counter-circular precessions around the $a$-axis. All measurements presented in this report are taken from a sample with its surface normal along the $a$-axis. We make this choice to optimize the signal from the M$_2$ mode and highlight the unique magnon properties in CTO.  

\begin{figure}
    \centering
    \includegraphics[width = 16 cm ]{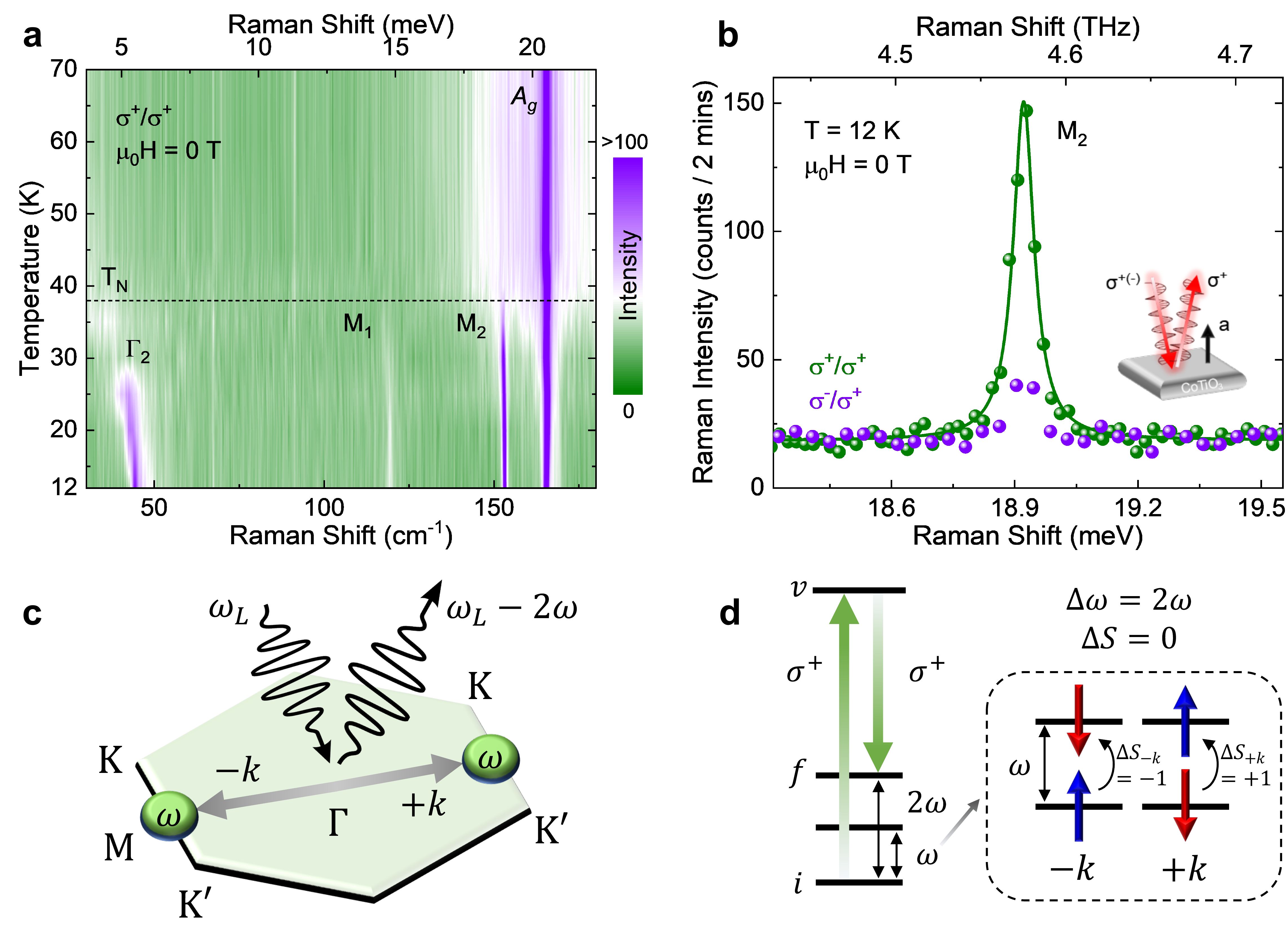}
    \caption{\textbf{Temperature and polarization dependence of Raman spectra identifying magnons and phonons.} (a) Temperature dependent Raman spectra taken at zero field. Multiple magnon modes at high symmetry points ($\Gamma_2$, M$_1$, M$_2$) are identified in the AFM phase, while a phonon mode ($A_g$) is observable in both the AFM and PM phases. (b) Polarization-resolved Raman spectra taken at 12 K in the AFM phase. $\sigma^+/\sigma^+$ (green dots) and  $\sigma^-/\sigma^+$ (purple dots) represent co- and cross-circularly polarized incident and scattered lights. Green solid line is a Lorentizian fit. Inset illustrates the back scattering geometry from a sample with surface normal along  the crystalline $a$-axis. (c) Schematic of the Stokes Raman process for two-magnon excitations. (d) Double spin-flip process in the $\sigma^+/\sigma^+$ channel. Two magnons ($2\omega$) with opposite wavevectors $+k$ or $-k$ are optically created via a virtual state and each undergoes a spin-flip. The net angular momentum change $\Delta S$ is $0$, leading to 2M signal in the $\sigma^+/\sigma^+$ polarization channel. ($i$: initial state; $f$: final state; $v$: virtual state.) 
    }
    \label{fig:fig2}
\end{figure}

\begin{figure}[H]
    \centering
    \includegraphics[width = 16cm ]{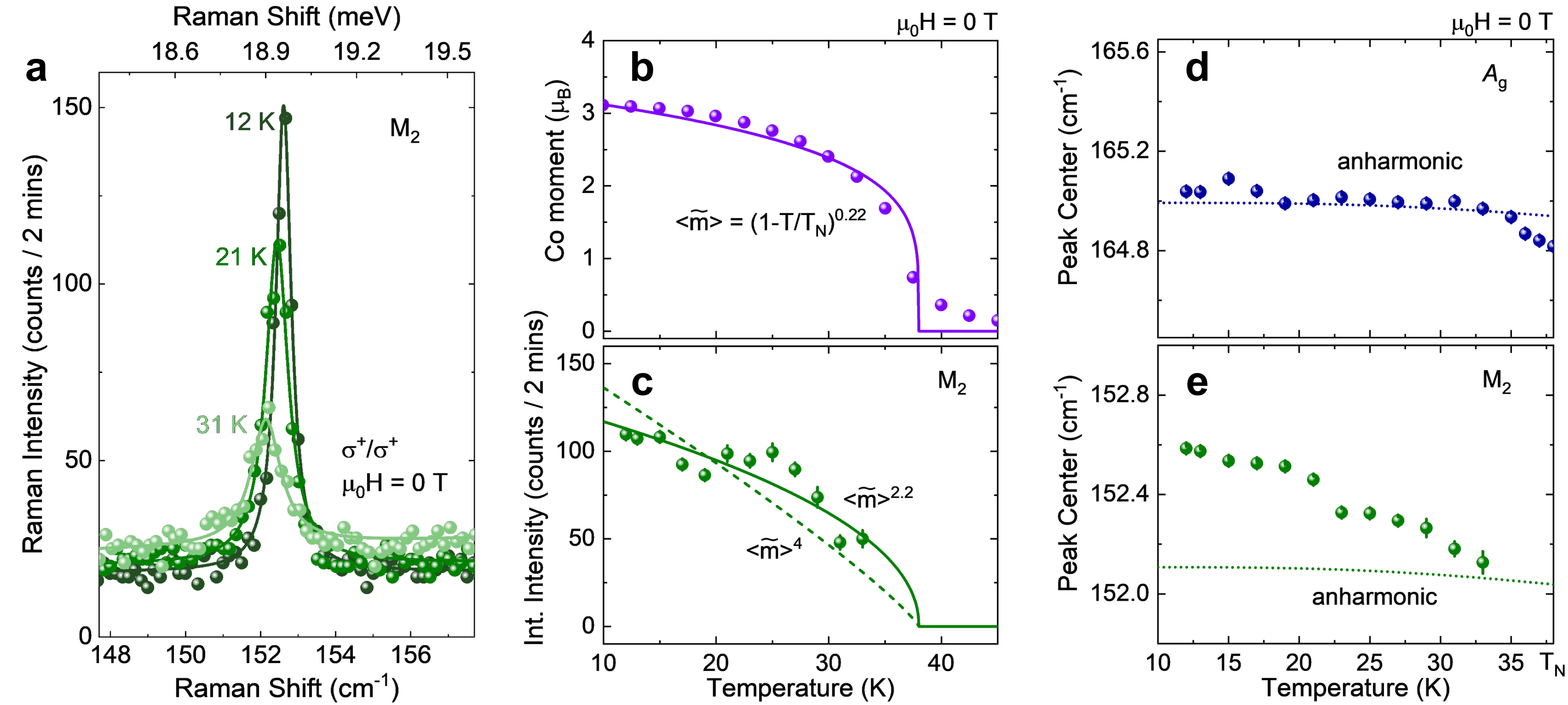}
    \caption{\textbf{Temperature dependence of M$_2$, Co magnetic moment, and $A_g$ phonon.} (a) M$_2$ Raman spectra taken at 12 K, 21 K, and 31 K at zero field in the $\sigma^{+}/\sigma^{+}$ channel. (b) Co magnetic moment extracted from neutron diffraction data from Ref.~\cite{elliot2021order}. Reduced sublattice magnetization $\langle\tilde{m}\rangle$ is fitted to an order parameter-like function $(1-T/T_N)^\beta$ with $\beta$ = 0.22 (purple solid curve). (c) Integrated intensity of M$_2$. The green solid curve $\langle\tilde{m}\rangle^{2.2 \pm 0.4}$ is the best fit through the measured Raman integrated intensity. Data deviates from the function $\langle\tilde{m}\rangle^{4}$ (green dashed line) predicted and observed for zone-folded phonons. Peak center of (d) $A_g$ phonon and (e) M$_2$ as a function of temperature. The M$_2$ frequency shift $\sim$ 0.5 cm$^{-1}$ exceeds that expected for a zone-folded phonon mode. Dashed lines in (d) and (e) are predicted frequency shifts from an anharmonic phonon model.}

    \label{fig:fig3}
\end{figure}

To identify magnons, we compare Stokes Raman spectra taken in two different magnetic phases. In the paramagnetic (PM) phase (T~$>$~T$_N$~$\sim$~38~K), a Raman-active phonon peak ($A_g$) is identified within the spectral range (Fig.~\ref{fig:fig2}a). In the AFM phase (T~$<$~T$_N$), multiple magnon peaks ($\Gamma_2$, M$_1$, M$_2$) are observed (more details shown in Supplementary Fig.~1). The energies, as well as the temperature and polarization dependence, of single magnon resonances agree
with those measured in INS experiments,
validating our comparison between Raman and INS experiments on CTO.  
We further attribute M$_{1}$ and M$_{2}$ peaks to two-magnon resonances dominated by the high symmetry M points on the Brillouin zone boundary (M$_{1}$, M$_{2}$) based on their frequencies in comparison to INS (Fig.~\ref{fig:fig1}b), the calculated high density of states (Fig.~\ref{fig:fig1}c), and temperature-dependent integrated intensity and peak center (Fig.~\ref{fig:fig3}) as further explained below.  Because of its strong intensity, we focus on the analysis of M$_{2}$ mode (Fig.~\ref{fig:fig2}b) in the following. Similarly, we assign the phonon modes based on temperature-dependent and polarization-resolved spectra, which are consistent with prior studies of phonons in CTO~\cite{DUBROVIN2020}.

In two-magnon Stokes scattering processes, incident photons ($\omega_L$) create a pair of magnons ($\omega$) so that the frequency of the scattered photon is shifted by twice the magnon frequency ($2\omega$). The pair of magnons have wavevectors and angular momenta equal in magnitude but opposite in direction as illustrated in Fig.~\ref{fig:fig2}c,d. Because the sum of the wave vectors is zero, momentum conservation is satisfied in the light scattering processes where photons carry negligible momentum. In principle, any point in the Brillouin zone can contribute to two-magnon scattering. The contribution from the zone-boundary magnons is enhanced by both a geometrical factor in the scattering cross-section~\cite{lockwood1987light} and their unique long lifetimes in CTO. Other zone-boundary magnons (e.g. K points) should also contribute. The higher density of states and its frequency closely matching the INS experiments suggest that the two-magnon resonance primarily originates from M-point magnons. Half of the $M_2$ magnon energy measured in Raman (green circles) is slightly lower than the single magnon energy measured in INS (purple circles) in Fig.~\ref{fig:fig1}b.

The angular momentum conservation requirement is typically manifested by the polarization selection rules. We qualitatively describe the scattering process leading to the M$_2$ mode in the co-circularly polarized channel. This two-magnon scattering process occurs via an exchange mechanism where a pair of exchange-coupled ($J_2$) magnetic ions within two spin sub-lattices simultaneously flip their spins upon the virtual electronic excitation as illustrated in Fig.~\ref{fig:fig2}d~\cite{fleury1968scattering}. Consequently, the total angular momentum change ($\Delta S$) is zero, thus, leading to a Raman signal in the  co-circularly polarized channel ($\sigma^+/\sigma^+$) as shown in Fig.~\ref{fig:fig2}b. 
While the dominant Raman signal in the co-circularly polarized channel can be qualitatively explained by this simplified picture, we caution that the low symmetry of the magnetic space group in CTO $P_S \bar{1}$~\cite{elliot2021order} does not constrain the Raman tensor enough to predict a strict selection rule via group theory.~\cite{yang2022signatures}

In addition to two-magnon resonances, zone-folded phonons can also emerge in the AFM phase below the N\'eel temperature. In order to rule out this alternative interpretation of the M$_{2}$ resonance, we analyze the temperature-dependent integrated Raman intensity and central frequency of the M$_{2}$ mode. We compare how the sublattice magnetization, M$_{2}$ and $A_g$ phonon mode evolve below the N\'eel temperature in Fig.~\ref{fig:fig3}. Raman spectra of M$_{2}$ taken at 30~K, 21~K, and 12 K are shown in Fig.~\ref{fig:fig3}a. With decreasing temperature, the M$_2$ peak exhibits an increasing integrated intensity and a blue shift in frequency as summarized in Figs.~\ref{fig:fig3}c and e. To explain these observations, we extract the Co sublattice magnetic moment $\langle m\rangle$ temperature dependence from neutron diffraction data~\cite{elliot2021order} in Fig.~\ref{fig:fig3}b. The magnetic moment $\langle m\rangle$ increases with decreasing temperature and saturates toward $\sim 3 \mu_B$. The reduced magnetic moment $\langle\tilde{m}\rangle=\langle m\rangle/3\mu_B$ is fitted to an order parameter-like function $(1-T/T_N)^\beta$, where $T_N = 38 K$ and $\beta = 0.22 \pm 0.02$ (also Supplementary Fig.~9). A two-magnon resonance is expected to scale with $\langle\tilde{m}\rangle^2$ while a zone-folded phonon mode scales with $\langle\tilde{m}\rangle^4$ as shown in CoPS$_3$ and FePS$_3$ previously~\cite{cottam1972theory,liu2021magnetic,scagliotti1987raman}. The fitting for the temperature-dependent M$_{2}$ Raman intensity yields $\langle\tilde{m}\rangle^{2.2 \pm 0.4}$. In Figs.~\ref{fig:fig3}c and e, we plot calculated intensity and frequency shift (dashed lines) if M$_{2}$ were a zone-folded phonon. The deviations between the measurements and phonon-based models support our assignment of M$_{2}$ to a two-magnon resonance.  We further identify the 228 cm$^{-1}$ peak as a zone-folded phonon mode in CTO and confirm that its temperature-dependent integrated intensity follows the $\langle\tilde{m}\rangle^4$ prediction (Supplementary Fig.~3).  

Next, we analyzed the temperature-dependent peak center shift for the nearby $A_g$ phonon (Fig.~\ref{fig:fig3}d) and M$_2$ modes (Fig.~\ref{fig:fig3}e). Both phonon population and frequency are expected to change with temperature. However, for a high-frequency optical phonon, the anharmonic model~\cite{PhysRevB.28.1928} (see dashed line in Fig.~\ref{fig:fig3}d) predicts a near constant frequency ($<$ 0.1 cm$^{-1}$) due to the small variation in the population of thermal phonons over this low-temperature range. In contrast, the frequency of the M$_2$ mode shifts by $\sim$ 0.5 cm$^{-1}$ exceeding that expected of a phonon mode, again validating the assignment of M$_2$ as a two-magnon resonance. We note that the M$_1$ peak also follows the same trends (See Supplementary Fig.~2), confirming that the properties of both M$_1$ and M$_2$ follow those predicted for two-magnon resonances. 

\begin{figure}[H]
    \centering
    \includegraphics[width = \textwidth ]{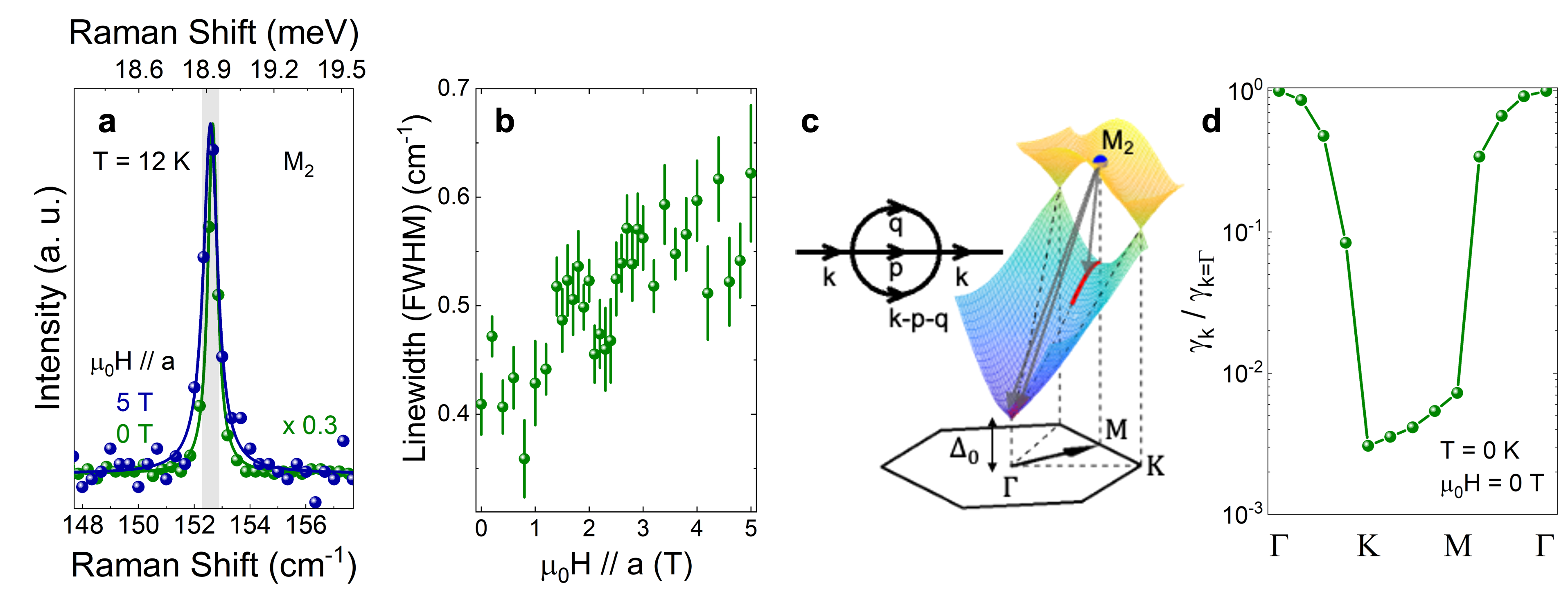}
    \caption{\textbf{Measured and calculated M$_2$ damping.} (a) M$_2$ Raman resonance measured at zero field (green) and 5 T (blue). 
    Solid lines are Lorentzian fits and the linewidth for the $\mu_0$H = 0 spectrum is limited by the instrument resolution, indicated by grey vertical strip. (b) Measured magnetic field dependence of M$_2$ linewidth extracted from Lorentzian fits. 
    (c) M$_2$ magnon decays to lower energy magnons (grey arrows) via a four-magnon process illustrated by the Feynman diagram. The lowest magnon energy is gapped ($\Delta_0$) at the $\Gamma$ point. As a result, four-magnon decay process is restricted (the red region) considering both the energy and momentum conservation. (d) Calculated magnon decay rate (normalized by $\Gamma$ point magnon) along a high-symmetry path at zero field and 0 K, showing $\sim140$ fold reduction for the zone-boundary magnons.}
    \label{fig:fig4}
\end{figure}

While two-magnon resonances are commonly observed in Raman spectra of AFM insulators~\cite{dietz1971infrared,tanabe2005direct,lujan2022magnons}, the narrow linewidth of the M$_2$ peak observed in CTO is remarkable. The magnon spectra shown in Fig.~\ref{fig:fig4}a are taken at zero field (green) and with a 5~T field (blue) applied along the $a$-axis. The M$_2$ linewidth at zero field is limited by the instrument resolution indicated by the grey vertical stripe. The linewidth broadens when the external magnetic field parallel to the $a$-axis, {\it i.e.} $\mu_0$H $\vert \vert$ $a$, increases as shown in Figs.~\ref{fig:fig4}b. This field-dependent line width can be partially attributed to additional decay channels arising from three-magnon interactions enabled in a non-collinear spin configuration (see Methods). A two-magnon resonance is spectrally broadened when either one of the two magnons with opposite momentum decays. Thus, half of the two-magnon resonance linewidth provides a lower bound on the single magnon lifetime $>$~25~ps.

To understand the origin of the long lifetime of zone-boundary magnons in CTO, we first examine their possible decay channels. In insulators, magnons primarily decay through two mechanisms: magnon-phonon and magnon-magnon interactions. Magnon and phonon coupling is strongest when their dispersions cross.~\cite{PhysRevB.49.4352,PhysRevB.42.4842,PhysRevLett.114.147201, mai2021magnon}. In CTO, however, both the acoustic and optical phonon modes have higher energies than magnons
at the zone-boundary, thus, reducing the magnon-phonon coupling~\cite{arruabarrena2022out}. Specifically, the M$_2$ and $A_g$ peaks are clearly separated at 12 K (see Fig.~\ref{fig:fig2}a). Both spectral line shapes are symmetric (Lorentzian), indicating the absence of quantum interference between magnon and phonon modes. We also perform temperature dependent measurements for the M$_2$ decay rate and identify a T$^{1.8}$ dependence, which supports magnon-magnon scattering~\cite{rezende1976multimagnon,kopietz1990magnon,nikitin2022thermal} as the dominant magnon decay channel rather than magnon-phonon scattering that should follow T$^5$.~\cite{cottam1974spin} (see Supplementary Fig.~12) Therefore, we focus on modeling magnon-magnon interactions below. The two-magnon continuum often overlaps with single-magnon bands at the zone boundary, leading to a rapid magnon decay. Thus, zone-boundary magnon modes typically exhibit a shorter lifetime than those at the zone center in common AFMs such as MnF$_2$ and NiO \cite{bayrakci2006spin,PhysRevB.93.054412}. Thus, the significantly narrower linewidth of zone boundary M$_2$ magnons than zone-center $\Gamma$ point magnons in CTO is a surprising observation.

To calculate the momentum-dependent magnon lifetime resulting from magnon-magnon interactions, we follow the procedure outlined below. This analysis not only provides insights into why other common AFM materials typically exhibit short-lived zone-boundary magnons but also establishes a framework for identifying long-lived zone-boundary magnons in other materials.  We begin with the general spin model $H =\sum_{i\delta} \textit{J}_1 (S^{x}_{i} S^{x}_{i+\delta} +S^{y}_{i}S^{y}_{i+\delta} + \alpha S^{z}_{i}S^{z}_{i+\delta}) + \sum_{i\gamma}\textit{J}_2 (S_{i} \cdot S_{i+\gamma})$ with exchange couplings $J_1$ = -4.4 meV and  $J_2$ = 0.57 meV. While the best and most precise Hamiltonian for CoTiO$_3$ is still under debate~\cite{elliot2021order,yuan2024field,li2024ring,choe2024magnetoelastic}, these models predict the same magnon dispersion measured by INS experiments. Thus, the choice among these Hamiltonians does not alter the calculated magnon lifetimes considering the assumptions adopted as described in Methods. For $\alpha = 0$, we recover the model for CTO, while for $\alpha=1$, we obtain an isotropic in-plane exchange interaction suitable for many van der Waals magnets such as CrI$_3$. Subsequently, we perform a Dyson-Maleev transformation to obtain the interacting magnon Hamiltonian, expressed as $H = H_{(0)}+H_{(2)}+H_{(3)}+H_{(4)}+\cdots$, where $H_{(0)}$ defines the ground state, $H_{(2)}$ is the non-interacting magnon Hamiltonian usually used to predict band topology, and $H_{(3)}$ ($H_{(4)}$) corresponds to the three (four)-magnon interaction (expressions are provided in the Methods section). These interaction terms are directly proportional to the magnetic exchange constants and give rise to finite magnon lifetimes. In the collinear AFM configuration at zero field, 3-magnon scattering is prohibited by spin-rotational symmetry around the N\`eel vector direction~\cite{RevModPhys.85.219}.

Assuming zero temperature and zero external magnetic field, the dominant decay channel for M$_2$ magnons corresponds to four-magnon decay processes in which one higher energy magnon spontaneously decays into three lower energy magnons as represented by the Feynman diagram in Fig.~\ref{fig:fig4}c. Multi-magnon decays arise from interactions between magnons with the bosonic character that the particle number is not conserved. Strikingly, our numerical evaluation of the momentum-dependent magnon lifetime predicts that the zone boundary magnon decay rate is lower by a factor of $\sim140$ than that for zone-center magnons in CTO as shown in Fig.~\ref{fig:fig4}d. We present the details of the momentum-dependent magnon lifetime calculations in Methods. We assume a constant parameter for matrix elements (see Methods) that may lead to a factor of 2 uncertainty in the predicted lifetimes. These calculations are meant to capture the key elements responsible for the momentum-dependent magnon lifetimes, rather than quantitative prediction of magnon lifetimes in CTO. The long-lived zone-boundary magnon in CTO can be partially attributed to a reduced joint two-magnon density of states (see Methods), consistent with a prior theoretical study on honeycomb-lattice XXZ AFMs~\cite{PhysRevB.93.014418}. The observed magnetic-field-dependent magnon linewidths (Fig.~\ref{fig:fig4}b) are readily explained by the additional three magnon scattering processes enabled in the non-collinear spin configuration in the presence of an external field. The three-magnon decay phase space is proportional to the two-magnon density of states, $D(\mathbf{q},E)=\frac{1}{N} \sum_{l,m}\sum_{\mathbf{p}}\delta(E-\varepsilon_{\mathbf{p},l}-\varepsilon_{\mathbf{q}-\mathbf{p},m})$. Our calculation confirms a larger number of decay channels at a finite magnetic field as shown in Supplementary Fig.~6. Although our experiments focused on M point magnons due to their higher density of states, Dirac magnons at K points may also exhibit longer lifetimes in CTO as shown in our calculated momentum-dependent lifetimes in Fig.~\ref{fig:fig4}d. Future experimental studies based on the neutron echo technique may confirm this prediction.

From a theoretical perspective, a honeycomb spin lattice with an interaction-induced magnon gap near the Dirac cone is expected to lead to topologically protected states. Among the materials proposed for topological magnons, three-magnon scattering is forbidden in AFMs generally and in selected FMs with spin-rotation symmetry.~\cite{RevModPhys.85.219} Additionally, those with well-separated phonon and magnon energy likely exhibit reduced phonon-magnon interaction-induced decay. CTO satisfies these simple principles. Our calculation taking into account the specific magnon dispersion of CTO also highlights that a shallow magnon dispersion combined with a sizable magnon gap ($\Delta_{0}$) at the $\Gamma$ point restricts the kinetic mechanism for magnon-magnon interaction-induced decay. 
Furthermore, exchange anisotropy ($\alpha=0$ in the generalized model) is found to be an unexpected yet critical element for long magnon lifetimes.  Empirically, a few families of honeycomb Van der Waals magnets including CrGe(Si)Te$_3$, CrX$_3$ (X=Cl, Br, I), and MPS$_3$ (M=Mn, Fe, Ni, Co)~\cite{chen2018topological, do2022gaps, cai2021topological, zhu2021topological,lee2018magnonic,bazazzadeh2021magnetoelastic} have been proposed or confirmed to host topological magnons but they all exhibit isotropic exchange interaction. We provide a survey of these alternative topological magnon materials in Supplementary Table~1. By calculating momentum-dependent magnon lifetimes in both CTO and CrI$_3$ (Methods and Supplementary Fig.~7 and 14), our studies suggest that a strong exchange anisotropy (XXZ-like magnon Hamiltonian) may be another critical element for long-lived zone-boundary magnons that has been largely ignored in previous studies. 


In summary, we discovered unusual magnon properties at the high symmetry points along the Brillouin zone boundary in CTO that hosts Dirac magnon bands. Because the saddle points in the magnon dispersion exhibit a large DOS, efficient two-magnon scattering processes allow us to observe large momentum magnons with nanometer wavelength. In contrast to other common AFMs, the zone-boundary magnons exhibit significantly longer lifetimes than those at the zone center. Our calculations explain the observed momentum- and field dependent magnon linewidths.  While we did not detect the elusive surface or edge states protected by topology, our study articulates several elements necessary to support long-lived magnons including AFM order, the small joint two-magnon density of states, exchange anisotropy, and single magnon bands. A long magnon lifetime is a necessary condition for both the detection of and harvesting the benefits of topologically protected states. By going beyond the single-particle topology and taking into account higher-order terms in the spin Hamiltonian, future computational searches can help to identify promising candidate materials for room-temperature topological magnons.~\cite{karaki2023efficient} 

\section*{Methods}
\subsection{Sample growth and characterization} 
High-quality CoTiO$_3$ single crystals were prepared using the floating zone method. Starting ceramic rods of CoTiO$_3$ were prepared by solid state reaction of thoroughly mixed stoichiometric of CoO (99.99\%, Alfa Aesar) and TiO$_2$ (99.99\%, Alfa Aesar) at 1000 $^{\circ}$C in air. The crystal growth in the image furnace was carried out with a constant air flow and 8 mm/h of growth speed. CoTiO$_3$ single crystal is black in color and the surface is shiny. The phase purity of CoTiO$_3$ was confirmed by powder X-Ray diffraction with the powder sample made by crushed single crystals. The sample orientation was determined by Laue back reflection method.\\

\subsection{Raman spectroscopy}
Raman measurements of CTO samples were performed using a 632.81 nm excitation laser in the back scattering geometry. The laser power of 0.1 mW was used to minimize the heating effect. The scattered light was collected for 2-min integration time, dispersed by an 1800-mm$^{-1}$ groove-density grating and detected by a thermoelectric-cooled CCD. The Raman shift difference between adjacent pixels of CCD was 0.34 cm$^{-1}$. The samples were mounted in a croystat capable of reaching temperature down to 10 K. A superconducting magnet was used to apply out-of-plane magnetic fields up to 5 T. \\

\subsection{Magnon dispersion calculations} 
The XXZ Hamiltonian with anisotropy exchange coupling is transformed by applying the Holstein-Primakoff transformation and reduced to $H=\sum_{\boldsymbol{k}} \mathbf{\Phi}_{\boldsymbol{k}}^{\dagger} \mathcal{D}(\boldsymbol{k}) \boldsymbol{\Phi}_{\boldsymbol{k}}$, where $\mathbf{\Phi}_{k}^{\dagger}=\left(a_{k}^{\dagger}, b_{k}^{\dagger}, c_{k}^{\dagger}, d_{k}^{\dagger}, a_{-k}, b_{-k}, c_{-k}, d_{-k}\right)$ and each of the operators $a,b,c,d$ annihilate magnons with momentum $k$ in the corresponding sublattice. The matrix $\mathcal{D}(\boldsymbol{k})$ is given by
\begin{equation}
\mathcal{D}(\boldsymbol{k})=\left(\begin{array}{cc}
\boldsymbol{A}(\boldsymbol{k}) & \boldsymbol{B}(\boldsymbol{k}) \\
\boldsymbol{B}^{\dagger}(\boldsymbol{k}) & \boldsymbol{A}^{*}(-\boldsymbol{k})
\end{array}\right)
\end{equation}
where, the matrices $\boldsymbol{A}(\boldsymbol{k}) $ and $\boldsymbol{B}(\boldsymbol{k})$ are given by

\begin{align}
\boldsymbol{A}(\boldsymbol{k}) & = \begin{pmatrix}
-z_1 S J_1+z_{\perp} S J_2 & S J_1\left(1+\frac{\alpha}{2}\right) f_{\boldsymbol{k}}   & 0 & 0 \\
S J_1\left(1+\frac{\alpha}{2}\right) f^*_{\boldsymbol{k}}  & -z_1 S J_1+z_{\perp} S J_2 & 0 & 0 \\
0  & 0 & -z_1 S J_1+z_{\perp} S J_2 & S J_1\left(1+\frac{\alpha}{2}\right) f_{\boldsymbol{k}} \\
0  & 0 & S J_1\left(1+\frac{\alpha}{2}\right) f^*_{\boldsymbol{k}} & -z_1 S J_1 +z_{\perp} S J_2
\end{pmatrix},
\end{align}

\begin{align}
\boldsymbol{B}(\boldsymbol{k}) & = \begin{pmatrix}
0   & S J_1 \frac{\alpha}{2}f_{\boldsymbol{k}}  & S J_2 f_{ac\perp,-\boldsymbol{k}} & S J_2 f_{ad\perp,-\boldsymbol{k}}\\
S J_1 \frac{\alpha}{2}f^*_{\boldsymbol{k}}  & 0 & S J_2 f_{bc\perp,-\boldsymbol{k}} & S J_2 f_{bd\perp,-\boldsymbol{k}} \\
S J_2 f^*_{ac\perp,-\boldsymbol{k}}  & S J_2 f^*_{bc\perp,-\boldsymbol{k}}  & 0 & S J_1 \frac{\alpha}{2}f_{\boldsymbol{k}} \\
S J_2 f^*_{ad\perp,-\boldsymbol{k}}  & S J_2 f^*_{bd\perp,-\boldsymbol{k}} & S J_1 \frac{\alpha}{2}f^*_{\boldsymbol{k}} & 0 
\end{pmatrix},    
\end{align}

$f_{\mathbf{k}} = \sum_{\boldsymbol \delta} e^{i  \mathbf k \cdot \boldsymbol{\delta}}$, and $z_1 = 3$ are the number of intra-plane nearest-neighbors, $f_{ac\perp,\mathbf{k}} = f_{bd\perp,\mathbf{k}} = \sum_{\boldsymbol \gamma} e^{i  \mathbf k \cdot \boldsymbol{\gamma}}$, $f_{ad\perp,\mathbf{k}} = \sum_{\boldsymbol \gamma} e^{i  \mathbf k \cdot \boldsymbol{\gamma_-}}$, $f_{bc\perp,\mathbf{k}} = \sum_{\boldsymbol \gamma} e^{i  \mathbf k \cdot \boldsymbol{\gamma_+}}$, and $z_{\perp}$ is the number of interlayer A-C next-nearest-neighbors. $\boldsymbol \gamma_{\mp}$ correspond to the interlayer nearest-neighbor vectors for A-D sublattices ($-$ sign) and B-C ($+$ sign) sublattices. The magnon energies are obtained by diagonalizing $\hat g \mathcal{D}(\boldsymbol{k})$ numerically, where $\hat g = [ 1 , 1 , 1, 1, -1 , -1 , -1, -1]$~\cite{COLPA1978327}. The presence of a spin-wave gap at the $\Gamma$ point lies beyond the XXZ model, and its origin is still under debate\cite{elliot2021order}. Here, we add a phenomenological term of the form $H_{\eta} = \frac{\eta}{2} \sum_{i,\boldsymbol \delta} \left( \tilde{S}^y_{i} \tilde{S}^y_{i+\boldsymbol \delta} - \tilde{S}^x_{i} \tilde{S}^x_{i+\boldsymbol \delta} \right)$, which leads to a gap at the $\Gamma$ point. We used $\eta = 0.05$~meV, which reproduces the low magnon energy of $\sim 1$~meV observed in neutron scattering measurements.~\cite{Yuan2020,elliot2021order}    \\

\subsection{Magnon lifetime calculations} 
In this section, we discuss general aspects of the magnon lifetime, arising from magnon-magnon interactions. To obtain the interaction potential, we perform a Dyson-Maleev transformation \cite{PhysRevB.3.961} to the spin Hamiltonian. The interacting magnon Hamiltonian can be written as $H = H_{(0)}+H_{(2)}+H_{(3)}+H_{(4)}+\cdots$. Here, $H_{(0)}$ corresponds to the classical energy and defines the ground state\cite{RevModPhys.85.219}. $H_{(2)}$ is the non-interacting magnon Hamiltonian leading to the magnon bands as discussed in the previous sections. $H_{(3)}$ and $H_{(4)}$ correspond to the three- and four-magnon interaction terms. 

We start with the zero-field case. We find $H_{(3)} = 0$, and
\begin{align*}
H_{(4)} & =\Biggl[\sum_{1234}[ J_1 \delta(1-2+3-4)f_{4-3}a^\dagger_1 a_2 b^\dagger_3 b_4 
+\frac{J_1}{2}\delta(1+2-3-4)f_{-2+3+4}\left(-1 +\frac{\alpha}{2}  \right)a^\dagger_1 b^\dagger_2 b_3 b_4 \\
&+\frac{J_1}{2}\delta(1-2-3+4)f_{-4}\left(-1 +\frac{\alpha}{2}  \right)a^\dagger_1 a_2 a_3 b^\dagger_4+\frac{J_1 \alpha}{4} \delta(1-2-3-4)f_{4} a^\dagger_1 a_2 a_3 b_4\\
&+\frac{J_1 \alpha}{4} \delta(-1+2-3-4)f_{-2+3+4} a_1 b^\dagger_2 b_3 b_4] +
        [(a,b) \rightarrow (c,d)]\Biggr]\\
-& \Biggl[\frac{J_2}{2}\sum_{1234}[  \delta(1-2-3-4)g_{4} a^\dagger_1 a_2 a_3 c_4    + 2 \delta(1-2+3-4)g_{-3+4} a^\dagger_1 a_2 c^\dagger_3 c_4\\
&+\delta(1+2+3-4)g_{-2-3+4} a^\dagger_1 c^\dagger_2 c^\dagger_3 c_4]\\&+[(a,c,\gamma) \rightarrow (b,d,\gamma)+(a,c,\gamma) \rightarrow (a,d,\gamma_-)+(a,c,\gamma) \rightarrow (b,c,\gamma_+)]\Biggr] 
\end{align*}
where $f_{\mathbf{k}} = \sum_{\boldsymbol \delta} e^{i  \mathbf k \cdot \boldsymbol{\delta}}$, $g_{\mathbf{k}} = \sum_{\boldsymbol \gamma} e^{i  \mathbf k \cdot \boldsymbol{\gamma}}$, and $\delta$ is the intralayer nearest-neighbor distance and $\gamma$ the interlayer nearest-neighbor distance between sublattices $a,c$; $\boldsymbol{\gamma}_-$ is between $a,d$; and  $\boldsymbol{\gamma}_+$ is between $b,c$. The numbers $1,2,3,4$ label the momenta $\boldsymbol k_1, \boldsymbol k_2, \boldsymbol k_3, \boldsymbol k_4$, respectively. The expression for $H_{(4)}$ contains terms that correspond to
magnon-conserving two-magnons-in$/$two-magnons-out processes (such as $a^\dagger_1 a_2 b^\dagger_3 b_4$), and decay and recombination of one magnon into three magnons (such as $a^\dagger_1 a_2 a_3 b_4$). The later terms that do not conserve the number of magnons are allowed in antiferromagnets because the ground state is a superposition of states with different total spin~\cite{anderson1984,PhysRevB.3.961,RevModPhys.85.219}. Diagrammatic representations of these terms are shown in Supplementary Fig.~4(a)~\cite{RevModPhys.85.219} and (b)~\cite{PhysRevX.8.011010}. 

The decay and recombination of one magnon into three magnons (Supplementary Fig.~4a) leads to a magnon lifetime of the form (at zero temperature) \cite{RevModPhys.85.219}
\begin{equation}
\Gamma_{\mathbf{k}}^{(4a)}=\frac{\pi}{6 N^2} \sum_{\mathbf{q},\mathbf{p}}\left|V_{4a}(\mathbf{q}, \mathbf{p} ; \mathbf{k})\right|^{2} \delta\left(\varepsilon_{\mathbf{k}}-\varepsilon_{\mathbf{q}}-\varepsilon_{\mathbf{p}}-\varepsilon_{\mathbf{k}-\mathbf{q}-\mathbf{p}}\right),
\label{eq:gammaa}
\end{equation}
where $\varepsilon$ are the magnon energies, and $V_{4a}(\mathbf{q}, \mathbf{p} ; \mathbf{k})$ is the momentum-dependent matrix element.  

The two-magnons-in/two-magnons-out scattering processes (Supplementary Fig.~4b) lead to a contribution to the lifetime~\cite{PhysRevX.8.011010}
\begin{equation}
\Gamma_{\mathbf{k}}^{(4b)}=\frac{\pi}{6 N^2} \sum_{\mathbf{q},\mathbf{p}}\left|V_{4b}(\mathbf{q}, \mathbf{p} ; \mathbf{k})\right|^{2} F(\mathbf{q}, \mathbf{p} ; \mathbf{k}) \delta\left(\varepsilon_{\mathbf{k}}+\varepsilon_{\mathbf{q}}-\varepsilon_{\mathbf{p}}-\varepsilon_{\mathbf{k}+\mathbf{q}-\mathbf{p}}\right),
\label{eq:gammab}
\end{equation}
where $F(\mathbf{q}, \mathbf{p} ; \mathbf{k})=\left[1+f\left(\varepsilon_{\mathbf{p}}\right)\right]\left[1+f\left(\varepsilon_{\mathbf{k}+\mathbf{q}-\mathbf{p}}\right)\right] f\left(\varepsilon_{\mathbf{q}}\right)-f\left(\varepsilon_{\mathbf{p}}\right) f\left(\varepsilon_{\mathbf{k}+\mathbf{q}-\mathbf{p}}\right)\left[1+f\left(\varepsilon_{\mathbf{q}}\right)\right]$, $f\left(\varepsilon_{\mathbf{q}}\right)=1 /\left[\exp \left(\varepsilon_{\mathbf{q}} / T\right)-1\right]$ accounts for its temperature dependence. 

Assuming that the matrix elements are a smooth function
of momentum, we approximate them with their average over momentum, $V_{4a(b)}(\mathbf{q}, \mathbf{p} ; \mathbf{k}) \rightarrow \langle V_{4a(b)} \rangle$ in Eqns. \eqref{eq:gammaa} and \eqref{eq:gammab}. We label the magnons bands $m_1(\boldsymbol k), m_2(\boldsymbol k), m_3(\boldsymbol k), m_4(\boldsymbol k)$, where $m_1(\boldsymbol k)$ corresponds to the lowest and $m_4(\boldsymbol k)$ to the highest energy band. 

We now analyze the magnon inverse lifetime arising from spontaneous decay (Eqn. \eqref{eq:gammaa}) considering different possible decay channels. We numerically integrate Eqn. \eqref{eq:gammaa} with momentum-averaged matrix elements using a Monte Carlo approach with $1200^2$ $k$-points randomly chosen in the BZ. The energy-conserving delta function is broadened with a factor of $10^{-2}$ meV. We compute the inverse magnon lifetime arising from decaying processes $m_4(\boldsymbol k) \rightarrow m_i(\boldsymbol q) + m_j(\boldsymbol p) + m_l(\boldsymbol k-\boldsymbol p-\boldsymbol q)$ for $i,j,l=1,2,3,4$ and find that the dominant channels involve magnons in the two lowest energy magnon bands. This is a consequence of a reduced two-magnon density of states, as we discuss later in the text. The decay channel $m_4(\boldsymbol k) \rightarrow m_1(\boldsymbol q) + m_1(\boldsymbol p) + m_1(\boldsymbol k-\boldsymbol p-\boldsymbol q)$ is shown in Figure 4d in the main text. This decay channel is the dominant one for temperatures much lower than the scale set by the magnon gap, $T \approx 10.9$~K.

Next, we consider the magnon spectral broadening due to thermal decays (e.g., two-magnons-in and two-magnons-out processes (Eqn. \eqref{eq:gammab}), with momentum-averaged matrix elements. We use a Monte Carlo approach with $1600^2$ $k$-points randomly chosen in the BZ. Since at $12$~K only the magnon band $m_1$ is thermally occupied near the $\Gamma$ point, the dominant decay channel for a $m_4(\boldsymbol k)$ magnon is of the form $m_4(\boldsymbol k) + m_1(\boldsymbol q\approx \Gamma) \rightarrow m_i(\boldsymbol p) + m_j(\boldsymbol k-\boldsymbol p-\boldsymbol q)$, with $i,j = 1, 2$. In Supplementary Fig.~13, we plot the inverse lifetime arising from the decay channel $m_4(\boldsymbol k) + m_1(\boldsymbol q\approx \Gamma) \rightarrow m_1(\boldsymbol p) + m_1(\boldsymbol k-\boldsymbol p-\boldsymbol q)$ for a set of temperatures up to $20$~K. We assume $\langle V_{4b} \rangle = \alpha \langle V_{4a} \rangle$, with $\alpha = 1$. However, a different $\alpha$ would scale the magnon spectral broadening due to thermal decay processes. The experimental measurements indicate $\alpha < 1$.

In the magnon lifetime calculations, from equations (4) and (5), we additionally assumed that the matrix elements are independent of the magnon wavefunctions and were approximated with a constant parameter. This simplification (in addition to the general assumptions outlined above) could lead to a factor of 2 or more differences in the predicted linewidth. The theory is not intended for a quantitative agreement but to identify the leading qualitative factors. We expect more sophisticated calculations will follow our experimental work.

For finite magnetic fields $H_z \parallel a$, we find the three-magnon interactions
\begin{align*}
H_{(3)} & =\sum_{123}[ \frac{g\mu_B H_z}{2\sqrt{2S}} \left(\cos \theta - i \sin \theta\right) \delta(1-2-3) (a^\dagger_1 a_2 a_3 + b^\dagger_1 b_2 b_3)] +[ (a,b,\theta) \rightarrow (c,d,-\theta)  ],
\end{align*}
where $\theta$ is the canting angle. The presence of additional decay channels from three-magnon interactions indicates that shorter lifetimes are naturally expected in non-colinear spin configuration, compared with the colinear spin. The four-magnon potential takes the form

\begin{align*}
H_{(4)} & =J_1(\sum_{1234}[ \delta(1-2+3-4)f_{4-3}a^\dagger_1 a_2 b^\dagger_3 b_4 -
\frac{1}{2}\delta(1-2-3+4)f_{-4}a^\dagger_1 a_2 a_3 b^\dagger_4\\
&-\frac{1}{2}\delta(1+2-3-4)f_{-2+3+4}a^\dagger_1 b^\dagger_2 b_3 b_4 \\
- & \frac{\alpha}{4}\delta(1-2-3+4)f_{-4}a^\dagger_1 a_2 a_3 b^\dagger_4-\frac{\alpha}{4} \delta(1+2-3-4)f_{-2+3+4}a^\dagger_1 b^\dagger_2 b_3 b_4\\
- & \frac{\alpha}{4}\left(\cos^2\theta - \sin^2\theta \right)(\delta(1-2-3-4)f_{4}a^\dagger_1 a_2 a_3 b_4
+  \delta(-1+2-3-4)f_{-2+3+4}a_1 b^\dagger_2 b_3 b_4 )\\
+& \frac{i \alpha}{2}\cos\theta \sin\theta (\delta(1-2-3-4)f_{4}a^\dagger_1 a_2 a_3 b_4+ \delta(-1+2-3-4)f_{-2+3+4}a_1 b^\dagger_2 b_3 b_4]\\
+&[ (a,b,\theta) \rightarrow (c,d,-\theta)  ])\\
-&(\frac{J_2}{2}\sum_{1234}[  \delta(1-2-3+4)g_{-4} a^\dagger_1 a_2 a_3 c^\dagger_4    - 2 \delta(1-2+3-4)g_{-3+4} a^\dagger_1 a_2 c^\dagger_3 c_4\\
&+\delta(1+2-3-4)g_{-2+3+4} a^\dagger_1 c^\dagger_2 c_3 c_4]\\&+[(a,c,\gamma) \rightarrow (b,d,\gamma)+(a,c,\gamma) \rightarrow (a,d,\gamma_-)+(a,c,\gamma) \rightarrow (b,c,\gamma_+)]). 
\end{align*}

The number of decay channels proportional to $J_2$ is the same as at zero field, but we obtained four more channels proportional to $J_1$ at the four-magnon interaction level. The structure of the expressions for the magnon lifetime is the same as at zero field. We note that additional terms could be present in the Hamiltonian due to the non-colinear magnetic moment orientation for finite applied magnetic field.

We also analyze the non-interacting two-magnon density of states, defined as~\cite{PhysRevLett.111.017204}
\begin{equation}
D(\mathbf{q},E)=\frac{1}{N} \sum_{l,m}\sum_{\mathbf{p}}\delta(E-\varepsilon_{\mathbf{p},l}-\varepsilon_{\mathbf{q}-\mathbf{p},m}),
\end{equation}
where $N$ is a normalization factor. $D(\mathbf{q},E)$ quantifies the phase space for the decay of two magnons. This is a relevant quantity because it appears explicitly in the magnon lifetime when we approximate the matrix element with its average over momentum $\langle V_4 \rangle$. For example, we can express the contribution from Supplementary Fig.~4a as $\Gamma_{\mathbf{k}}^{(4a)} \approx \frac{\pi}{6 N} \sum_{\mathbf{q};ij}\left|\langle V_4 \rangle\right|^{2} \int dE \delta(E-\varepsilon_{\mathbf{k},i}+\varepsilon_{\mathbf{q},j}) {D}(\mathbf{k}-\mathbf{q},E).$ In Supplementary Fig.~5, we plot $D(\mathbf{q},E)$ along a high-symmetry path in the BZ as a function of energy. At the boundary of the BZ, the two-magnon density of states is small compared with the $\Gamma$ point. 


We plot the non-interacting two-magnon density of states $D(\mathbf{q},E)$ for the magnon energies of M$_1$ and M$_2$ and compare between 0 T and 4.5 T (Supplementary Fig.~6). The number of magnon decay channels at zone boundary is highly suppressed for M$_1$ magnons, indicating longer magnon lifetime. M$_2$ magnons exhibit relatively smaller decay rates compared to $\Gamma$ point. Both M$_1$ and M$_2$ magnons show the increasing number of decay channel at the finite field.

To justify our model and provide a meaningful comparison to CTO, we calculated the magnon lifetime of CrI$_3$ at zero magnetic field, an extensively studied material with confirmed Dirac magnon dispersion (see Supplementary Fig.~14). As opposed to CTO, bulk CrI$_3$ is ferromagnetic with isotropic exchange interactions and a gapped Dirac dispersion at the K-point~\cite{chen2021magnetic}. We found that, despite the underlying hexagonal layered structure of the lattice, the M-point magnons are not significantly longer-lived than the $\Gamma$-point magnons, different from CTO.\\

\section*{Data availability}
The data that support the plots within this paper and other findings of this study are available from the corresponding authors upon reasonable request. Source data are provided with this paper.

\section*{References} 

\section*{Author Contributions}
J.H. and J.Z. grew the CTO bulk crystals and characterized the samples. J.C., D.L.,G.Y., and C.N. performed magneto-Raman measurements under the supervision of R.H. and X.L. Data were analyzed by J.C, D.L., F.Y.G., T.N.N., R.H., and X.L. The model calculations are performed by M.R.V., J.C., B.M. A.L., M.A., and A.A. J.C., G.F., M.R.V., R.H. and X.L. wrote the paper with the input from all authors. 

\section*{Acknowledgements} 
This research (J.C. and D. L.) was primarily supported by the National Science Foundation through the Center for Dynamics and Control of Materials: an NSF MRSEC under Cooperative Agreement No. DMR-2308817. X. Li gratefully acknowledges funding from the Center for Energy Efficient Magnonics an Energy Frontier Research Center funded by the U.S. Department of Energy, Office of Science, Basic Energy Sciences at SLAC National Laboratory under contract DE-AC02-76SF00515 (discussion of long-lived magnons for applications) and the Welch Foundation Chair F-0014 (materials supply). Additional support from NSF DMR-2114825, DOE Award DE-SC0022168, and the Alexander von Humboldt Foundation is gratefully acknowledged by G.A.F. This work was performed in part at the Aspen Center for Physics, which is supported by National Science Foundation grant PHY-1607611 and PHY-2210452 (X. Li). R.H. acknowledges support by NSF Grants No. DMR-2300640 and DMR-2104036 and DOE Office of Science Grant No. DE-SC0020334 subaward S6535A. Part of the experiments were performed at the user facility supported by the National Science Foundation through the Center for Dynamics and Control of Material under Cooperative Agreement No. DMR-1720595 and The Major Research Instrumentation (MRI) program DMR-2019130.\\

\section*{Competing Interests:} The authors declare no competing interests.\\

\newpage
\section*{Figures} 
\renewcommand{\baselinestretch}{1}

\end{document}